\documentclass[a4paper,12pt]{article}
\usepackage[pctex32]{graphics}

\begin{document}

\newcommand{\beq}{\begin{equation}}
\newcommand{\eeq}{\end{equation}}
\newcommand{\beqa}{\begin{eqnarray}}
\newcommand{\eeqa}{\end{eqnarray}}
\newcommand{\sr}{\sqrt}
\newcommand{\fr}{\frac}
\newcommand{\mn}{\mu \nu}
\newcommand{\G}{\Gamma}

\begin{titlepage}
\begin{flushright}
INJE-TP-06-04,~astro-ph/0605684
\end{flushright}

\vspace{5mm}
\begin{center}
{\Large \bf Gauss-Bonnet braneworld  and WMAP three year results}
\vspace{12mm}

{\large    Brian M. Murray$^{\rm a}$\footnote{e-mail
 address: bmurray1@uoregon.edu}
and Yun Soo Myung$^{\rm a,b}$ \footnote{e-mail
 address: ysmyung@inje.ac.kr}}
 \\
\vspace{10mm} {\em  $^{\rm a}$Institute of Theoretical Science,
University
of Oregon, Eugene, OR 97403-5203, USA \\
$^{\rm b}$Institute of Mathematical Sciences, Inje University,
Gimhae 621-749, Korea}
\end{center}
\vspace{5mm} \centerline{{\bf{Abstract}}}
 \vspace{5mm} We compare predictions for the spectral index and
tensor-scalar ratio in models of patch inflation with the WMAP
three year data. There are three cases of these models of
inflation, which arise in the Gauss-Bonnet braneworld scenario:
Gauss-Bonnet (GB), Randall-Sundrum (RS), and 4D general relativity
(GR). We consider the large-field potential $V \propto \phi^p$ in
both commutative and noncommutative spacetimes, and find that in
the cases of the GB and GR patch cosmologies, the quadratic
potential is observationally favored, while the quartic potential
is ruled out in most patches. Strong noncommutative inflation is
excluded in all cases because it leads to a blue-tilted scalar
spectral index.
\end{titlepage}
\newpage
\renewcommand{\thefootnote}{\arabic{footnote}}
\setcounter{footnote}{0} \setcounter{page}{2}

\section{Introduction}

In recent years there has been much interest in the phenomenon of
localization of gravity proposed by Randall and Sundrum
(RS)~\cite{RS2}. They assumed a three-brane with a positive
tension embedded in 5D anti-de Sitter (AdS$_5$) spacetime, and
were able to localize gravity on the brane by fine-tuning the
brane tension to the bulk cosmological constant. Recently, several
authors have studied the cosmological implications of braneworld
scenarios. As would be expected, brane cosmology contains some
important deviations from the Friedmann-Robertson-Walker
cosmology~\cite{BDL}.

At the same time, it is generally thought that curvature perturbations
produced during inflation may be the origin of the inhomogeneity that
is necessary to explain the anisotropy in the cosmic microwave
background as well as the presence of large-scale structure. The WMAP
first year results (WMAP1)~\cite{Wmap1}, SDSS~\cite{SDSS,SDSS2}, and
other data lead to more constraints on cosmological models. As a
result of combining these results from various observations, the ${\rm
\Lambda}$CDM model has emerged in recent years as the standard model
of cosmology. Moreover, these results coincide with the theoretical
prediction of slow-roll inflation with a single inflaton field.

Recently, the WMAP three year results (WMAP3)~\cite{WMAP3}
obtained values for the spectral index
$n_s=0.951^{+0.015}_{-0.019}~(0.948^{+0.015}_{-0.018})$ and the
tensor-scalar ratio $r<0.55~(0.28)$ for WMAP3 alone (WMAP3+SDSS)
at the $2\sigma$ level.  More recently, combined data including
WMAP3, SDSS, Lyman-$\alpha$, SN Ia and galaxy clustering (combined
DATA) indicated $n_s=0.965^{+0.012+0.025}_{-0.012-0.024}$ and
$r<0.22$~\cite{SSM}.  It would appear that a red power spectrum
with $n_s<1$ is favored, while a scale-invariant
Harrison-Zel'dovich-Peebles (HZ) spectrum ($n_s=1,r=0$) is
disfavored at the 2$\sigma$ level.  The authors of~\cite{KKMR},
however, reported that the HZ spectrum is consistent with both
WMAP3 and WMAP3+SDSS at the $2\sigma$ level.  We use the combined
DATA in this work due to the fact that the contours for WMAP3 and
WMAP3+SDSS in Fig. 14 of Ref.~\cite{WMAP3} were
incorrect~\cite{LEW,KKMR}. If one allows for a running spectral
index $\alpha_s$, the fit to the WMAP3 data is slightly improved.
The improvement, however, is not significant enough to require the
running. Hence, we choose to neglect running ($\alpha_s \simeq 0$)
for comparison with theoretical values. Importantly, it is shown
that chaotic inflation with $V(\phi) \sim \phi^2$ fits the
observations very well. It is certainly the case that the WMAP3
data with or without additional observations provides significant
constraints on models of inflation and some models are ruled out
at a high level of confidence~\cite{AL}.

If inflation occurs on the brane, one would expect that it
provides us quite different results in the high-energy
region~\cite{MWBH}.  Since the Gauss-Bonnet term significantly
modifies the Friedmann equation at high-energy, its application to
brane inflation has been studied widely in the
literature~\cite{DLMS}. Moreover, noncommutative spacetime
naturally emerges in string theory, and although a realistic
inflationary model has not yet been constructed in noncommutative
spacetime, there is a toy model of noncommutative inflation
~\cite{BH}.

In this work, patch cosmological models that arise in the
Gauss-Bonnet braneworld scenario are used to study brane inflation for
large-field potentials in noncommutative spacetime. We use the
leading-order theoretical predictions for the spectral index and
tensor-scalar ratio to determine which patch models are consistent
with the combined DATA.

The organization of this work is as follows. In section 2 we briefly
review patch cosmology in noncommutative spacetime. We introduce
large-field potentials, compute their theoretical values of $n_s$ and
$r$, and compare these predictions with the observation data in
section 3. Finally, we discuss our results in section 4.

\section{Patch cosmological models}

We start with the action for the  Gauss-Bonnet braneworld
scenario~\cite{DLMS}: \begin{eqnarray}
S&=&\fr{1}{2\kappa^2_5}\int_{{\rm
bulk}}d^5x\sqrt{-g_5}\Big[R-2\Lambda_5+\alpha\Big(R^2-4R_{\mu\nu}R^{\mu\nu}+
R_{\mu\nu\rho\sigma}R^{\mu\nu\rho\sigma}\Big)\Big] \nonumber \\
     &+&\int_{{\rm brane}}d^4x \sqrt{-g}\Big[-\lambda
+{\cal L}_{{\rm matter}} \Big],
\end{eqnarray}
where $\Lambda_5=-3\mu^2(2-\beta)$ is the AdS$_5$ bulk
cosmological constant, with the AdS$_5$ energy scale $\mu=1/\ell$.
${\cal L}_{{\rm matter}}$ is the matter lagrangian for the
inflaton field. $\kappa^{2}_{5}=8\pi/m_{5}^{3}$ is the 5D
gravitational coupling constant and $\kappa^2_4=8\pi/m_{{\rm
Pl}}^{2}$ is the 4D coupling constant. The Gauss-Bonnet coupling
$\alpha$ may be related to the string energy scale $g_s$ ($\alpha
\simeq 1/8g_s$) when the Gauss-Bonnet term is considered to be the
lowest-order stringy correction to the 5D gravity.  $\lambda$ is
the brane tension. Relations between these quantities are
$\kappa^2_4/\kappa^2_5=\mu/(1+\beta)$ and
$\lambda=2\mu(3-\beta)/\kappa_5^2$, where $\beta=4\alpha\mu^2\ll
1$. The RS case of $\mu=\kappa_4^2/\kappa_5^2$ is recovered for
$\beta=0$ $(\alpha=0)$. We have to distinguish between the GB
($\beta \ll1$, $\beta\not=0$) and RS ($\beta=0$) cases.  The exact
Friedmann-like equation is given by a complicated form,
\begin{equation} \label{Eeq}
2\mu\sqrt{1+\frac{H^2}{\mu^2}}\Big[3-\beta+2\beta
\frac{H^2}{\mu^2}\Big]=\kappa_5^2(\rho+\lambda),
\end{equation}
where as usual $H=\dot{a}/a$.
We, however, use  an effective  Friedmann equation
\begin{equation}
\label{Heq} H^2= \beta^{2}_{q}\rho^{q}.
\end{equation}
Here $q$ is a patch parameter labeling different models, and
$\beta_{q}^{2}$ is a corresponding factor with energy dimension
$[\beta_q]=E^{1-2q}$.  We call the above  ``patch cosmology," and
summarize the three different models and their parameters  in Table 1.

\begin{table}
 \caption{Three patch cosmological models and the values of the
parameters that classify them.
 Here $m^4_{\alpha}=[8 \mu^2(1-\beta)^3/\beta \kappa^4_5]^{1/2}$ is the GB energy scale.}
\begin{center}
 \begin{tabular}{|c|c|c|c|c|c|}
 \hline
 model   & $q~(\zeta_q)$ & $\beta^2_q$ & acceleration $(\omega_q$)& $\rho$  \\ \hline
 GB      & 2/3~(1)    & $(\kappa^2_5/16\alpha)^{2/3}$ & $-1 \le \omega<0$ & $\rho \gg m^4_{\alpha}$\\ \hline
 RS      & 2 ~(2/3)      & $\kappa^2_4/6\lambda$ & $-1 \le \omega<-2/3$& $\lambda \ll \rho \ll m^4_{\alpha}$\\ \hline
 GR      & 1 ~(1)        & $\kappa_4^2/3$ & $-1 \le \omega<-1/3$& $\rho \ll \lambda$ \\ \hline

 \end{tabular}
\end{center}
 \end{table}

 Before we proceed, we note that
the Gauss-Bonnet braneworld affects inflation only when the Hubble
parameter is much larger than the AdS  scale ($H\gg \mu$). As a
result, there are two patch models, the GB case with $q=2/3$ and
the RS case with $q=2$. For  $H \ll \mu$, one recovers the 4D
general relativistic (GR) case with $q=1$.

On the brane, let us introduce  an inflaton field $\phi$ whose equation
is given by
\begin{equation}
\label{seq} \ddot{\phi}+3H\dot{\phi}=-V^{\prime},
\end{equation}
where dot and prime denote  the derivative with respect to time
$t$ and $\phi$, respectively. The energy density and pressure are
given by $\rho=\dot{\phi}^2/2+V$ and $p=\dot{\phi}^2/2-V$. From
now on, we use the slow-roll formalism for inflation: an
accelerating universe $(\ddot a>0)$ is driven by an
inflaton slowly rolling down its potential toward a local minimum.
Then Eqs.~(\ref{Heq}) and (\ref{seq}) take the approximate form
\begin{equation}
H^2\approx \beta^2_q V^q,~ \dot{\phi}\approx -V'/3H .
\end{equation}
These are the equations for the background. In order for an inflation
to terminate and for the universe to transition to a
radiation-dominated phase, there must be a slow-roll mechanism. To
this end, it is conventional to introduce Hubble slow-roll parameters
($\epsilon,\delta$) and potential slow-roll parameters
($\epsilon^q,\delta^q$),
\begin{equation}\label{srp}
\epsilon = -\frac{\dot H}{H^2}\approx \epsilon^q =
\fr{q}{6\beta^2_q}\fr{(V')^2}{V^{1+q}},~~\delta =
\frac{1}{H\dot{\phi}}\frac{d^{2}\phi}{dt^{2}}\approx
\delta^q=\frac{1}{3\beta^2_q}\Big[\frac{q}{2}\frac{(V')^2}{V^{1+q}}-\frac{V''}{V^q}\Big].
\end{equation}
The slow-roll parameter $\epsilon^q\ge 0$ governs
the equation of state $p=\omega_q\rho$ with
$\omega_q=-1+2\epsilon^q/3q$. This implies that an accelerating
expansion occurs only for
$\epsilon^q<1~(\omega_q<-1+2/3q)$~\cite{Kinn}.
$\epsilon^q=0~(\omega_q=-1)$ corresponds to de Sitter inflation.
The end of inflation is determined by
$\epsilon^q=1~(\omega_q=-1+2/3q)$. Hence, the allowed regions for
acceleration (inflation) depend on $q$, as shown in Table 1.
If one chooses an inflation potential $V$, then potential
slow-roll parameters ($\epsilon^q,\delta^q$) are determined
explicitly.

Noncommutative inflation arises by imposing a realization of the
$*$-algebra on the brane coordinates: $[\tau,x]=i l_s^2$, where
$\tau=\int adt$, $x$ is a comoving spatial coordinate, and
$l_s=1/M_s$ is the  string scale\cite{BH,Li,CT,KLM}. By introducing
the noncommutative parameter $\delta=(M_s/H)^2$, the
noncommutative algebra induces a cutoff $k_0(\delta)$, which
divides the space of comoving wavenumbers into two regions: the
UV-commutative perturbations generated inside the Hubble horizon
($H \ll M_s$) and the IR-noncommutative perturbations generated
outside the horizon ($H \gg M_s$). In this case the amplitude of
scalar perturbations is given by~\cite{CT}
\begin{equation}
\label{amp}
A_s^2=\frac{9\beta^6_q}{25\pi^2}\frac{V^{3q}}{V^{\prime 2}}
\Sigma^2(\delta),
\end{equation}
where $\Sigma(\delta)$ is a function that includes noncommutative
effects. The above amplitude is evaluated at horizon crossing in the
UV-limit, and at the time when the perturbation with comoving
wavenumber $k$ is generated in the IR-limit.  To the lowest-order in
the slow-roll parameters, we have
\begin{equation}
\label{damp} \frac{d \ln \Sigma^2}{d \ln k}=\tilde{\sigma}
\epsilon^q ,
\end{equation}
where $\tilde{\sigma}=\tilde{\sigma}(\delta)$ is a function of
$\delta$ and  $\dot{\tilde{\sigma}}={\cal O}(\epsilon^q)$. Here we
choose three cases in the far IR-limit: $\tilde{\sigma}=0$
(UV-commutative case), $\tilde{\sigma}=2$ ($\Sigma^2\sim \delta$),
and $\tilde{\sigma}=6$ ($\Sigma^2\sim \delta^3$).
$\tilde{\sigma}=2~(6)$ correspond to the weak (strong)
noncommutative case.

\begin{figure}[t!]
 \centering
   \includegraphics{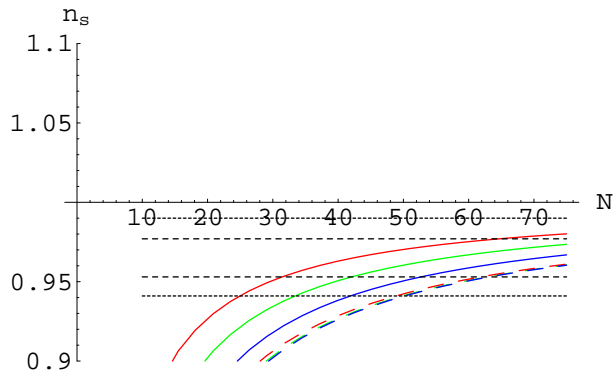}
    \caption{The spectral index $n_s$ versus number of e-foldings $N$
    for the commutative case $\tilde{\sigma}=0$.  The red, green, and
    blue curves represents the GB, GR, and RS cases,
    respectively. Solid (long-dashed) lines denote the case of
    $p=2~(4)$.  The short-dashed lines denote the 1$\sigma$ interval
    between 0.953 and 0.977, while the dotted lines show the 2$\sigma$
    interval between 0.941 and 0.990~\cite{SSM}.  } \label{fig1}
\end{figure}

\begin{figure}[t!]
 \centering
   \includegraphics{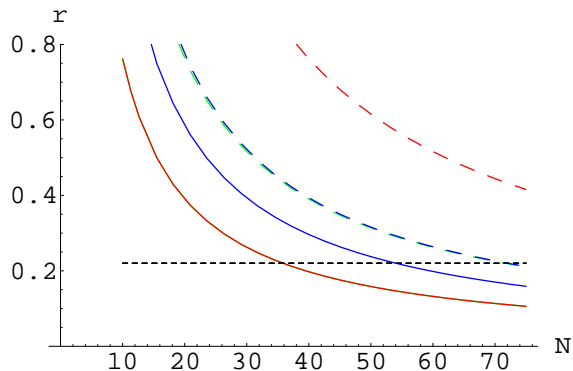}
    \caption{The tensor-scalar ratio $r$ versus number of e-foldings $N$,
     with parameter values and colorings as in Fig.~\ref{fig1}.
     The short-dashed line denotes
    the 2$\sigma$ level of $r<0.22$~\cite{SSM}.  } \label{fig2}
\end{figure}

The $q$-spectral index is
given by
\begin{equation}
\label{11} n^{q}_{s}(k) = 1 - (4-\tilde{\sigma})\epsilon^q -
2\delta^q,
\end{equation}
and the tensor-scalar ratio $r_q$ is given  by \beq \label{ttosr}
r_q=16\fr{\epsilon^q}{\zeta_q},\eeq where $\zeta_q$ is given by
Table 1. Note that $r_q$ is independent of the noncommutative
parameter.

\section{Inflation with large-field potentials}

\begin{figure}[t!]
 \centering
   \includegraphics{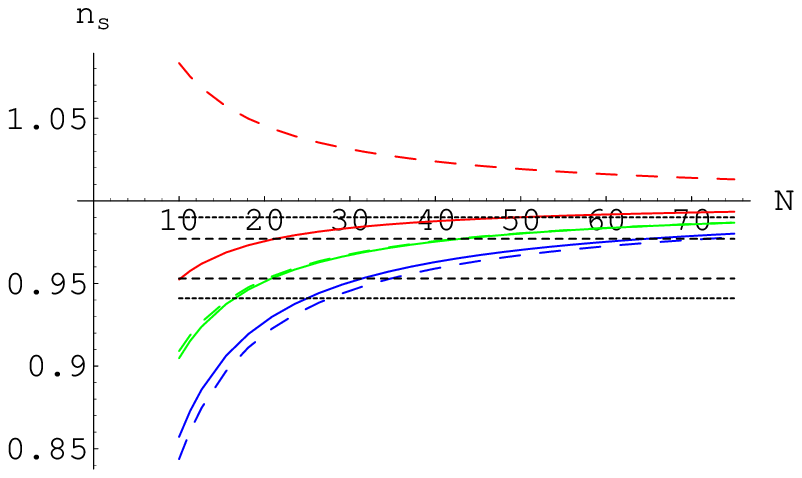}
    \caption{The spectral index $n_s$ versus the number of e-foldings
    $N$ for the weak noncommutative case $\tilde{\sigma}=2$. All other
    parameters, colors, and lines are as in Fig.~\ref{fig1}.}
    \label{fig3}
\end{figure}

\begin{figure}[t!]
 \centering
   \includegraphics{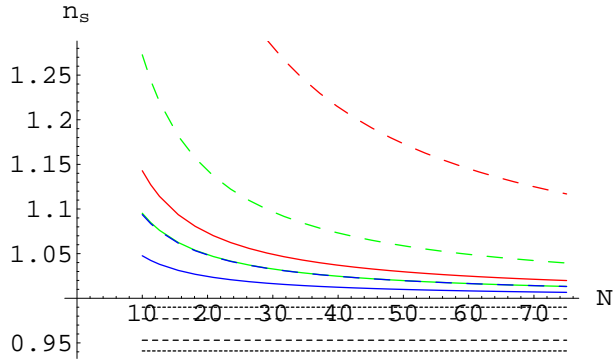}
    \caption{The spectral index $n_s$ versus the number of e-foldings
    $N$ for the strong noncommutative case $\tilde{\sigma}=6$. All other
    parameters, colors, and lines are as in Fig.~\ref{fig1}.}
    \label{fig4}
\end{figure}

A single-field potential can be characterized by two energy scales:
the height of potential $V_0$ corresponding to the vacuum energy
density for inflation and the width of the potential $w$ corresponding
to the change in the inflaton $\Delta \phi$ during inflation. In
general, the potential can be expressed as $V=V_0f(\phi/w)$. Different
potentials have different choices for the function $f$. The height
$V_0$ is usually fixed by normalization so that $w$ is the only free
parameter in the potential. Here we focus on the case of large-field
potentials, $V^{LF}=V_{0p}\phi^p$ for comparison with the combined
DATA. For $p=2$, $V_{02}=m^2/2$ is the case of a massive scalar, and
for $p=4$, $V_{04}=\lambda/4$ is the case of a model with a
self-coupling. In these cases, the potential slow-roll parameters are

\begin{eqnarray}\label{LFsl}
\epsilon^q_p &=& {{qp} \over 2} \frac{1}{X},\\
\delta^q_p &=& {1 \over 2}{(2 - 2p + qp)  \over X},
\end{eqnarray}
where we have defined $X \equiv  [(q-1)p + 2]N + {{qp} \over 2}$.
Substituting these expressions into Eqs.~(\ref{11}) and (\ref{ttosr}),
we obtain the desired results for $n_s$ and $r$.
In the leading-order calculation, the LF-spectral
index in noncommutative spacetime is given by
\begin{equation}\label{LFnq}
n_{s}^{LF} =1- \frac{(3q-q \tilde{\sigma}/2-2)p+2}{X}.
\end{equation}
The LF tensor-scalar ratio takes the form \label{LFRq}\beq
r^{LF}=\frac{8qp}{\zeta_q} \frac{1}{X}.  \eeq Fortunately, there
are no additional free parameters for large-field potentials, once
a choice has been made for the e-folding number $N$. See Figs. 1-4
for plots. In Fig. 1, three long-dashed curves are nearly the same
for $p=4$, although the three solid ones are distinctive for
$p=2$. In Fig. 2 consists only of monotonically decreasing
functions of $N$. For $p=2$, the solid GB and GR curves are
degenerate, while for $p=4$, the long-dashed GR and RS curves are
degenerate.  In Fig. 3 we plot $n_s$ for the weak noncommutative
case $\tilde{\sigma}=2$.  Here all curves are monotonically
increasing functions except the case of $p=4$ in GB. The curves of
GR for $p=2$ and $p=4$ are nearly the  same, and the two cases of
RS patch are even more degenerate.  Finally, in Fig. 4 we plot
$n_s$ for the strong noncommutative case $\tilde{\sigma}=6$. The
curves for $p=2$ in GR and $p=4$ in RS are degenerate.  Here all
curves are monotonically decreasing functions above the 2$\sigma$
level, which shows that the strong noncommutativity leads to the
blue-tilted scalar spectra with $n_s>1$ for large-field
potentials.

\begin{table}
\caption{The spectral index $n_s$ and tensor-scalar ratio $r$ for
three patch models (determined by $q$). We focus on the cases $N=50$
and $60$ to obtain theoretical values for large-field potentials
$V=V_{0p}\phi^p$. Here we indicate
$n_s=(\tilde{\sigma}=0,\tilde{\sigma}=2,\tilde{\sigma}=6)$. }
\begin{center}
 \begin{tabular}{|c|c|c|c|}
 \hline  Patch & $p$ & $N=50$& $N=60$ \\ \hline
       GB &2& $n_s=(0.970,0.990,1.030),r = 0.16 $ & $n_s=(0.975,0.992,1.025),r = 0.13 $ \\ \cline{2-4}
   $(q=2/3)$ &4& $n_s=(0.942,1.019,1.173),r = 0.62 $ & $n_s=(0.952,1.016,1.145),r = 0.52 $ \\ \hline

       RS &2& $n_s=(0.950,0.970,1.010),r = 0.24 $ & $n_s=(0.959,0.975,1.008),r = 0.20 $  \\
         \cline{2-4}
       $(q=2)$ &4& $n_s=(0.941,0.967,1.020),r = 0.32 $ & $n_s=(0.951,0.973,1.016),r = 0.26 $ \\
         \hline
       GR  &2& $n_s=(0.960,0.980,1.020),r = 0.16 $ & $n_s=(0.967,0.983,1.017),r = 0.13 $ \\ \cline{2-4}
    $(q=1)$ &4& $n_s=(0.941,0.980,1.059),r = 0.31 $ & $n_s=(0.951,0.984,1.049),r = 0.26 $ \\\hline

\hline
\end{tabular}
\end{center}
\end{table}

\section{Discussion}

\begin{figure}[t!]
  \centering
   \includegraphics{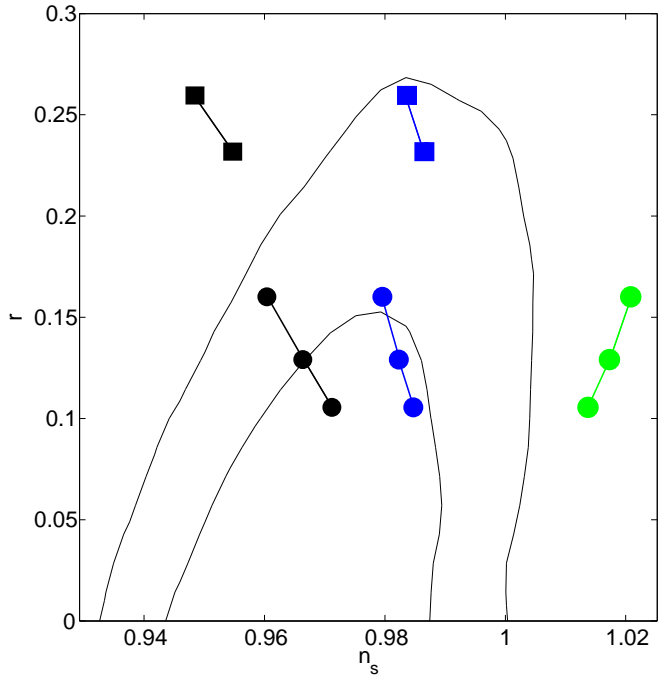}
    \caption{Observational constraints ($1\sigma~(68\%$ C.L.) and
    $2\sigma~(95\%$ C.L.) contours) in the $n_s$-$r$ plane on
    large-field potentials for the GR patch. The theoretical values
    correspond to (a) $p=2$ (dots) with $N$=50, 60 and 70 (top to bottom)
      Three classes of (non)commutative
    inflation are at left ($\tilde{\sigma}=0$:black), center
    ($\tilde{\sigma}=2$:blue), and right ($\tilde{\sigma}=6$:green). (b) $p=4$ (squares) with $N$=60 and 70.
    Here $\tilde{\sigma}=6$ is absent. }
    \label{fig5}
\end{figure}

\begin{figure}[t!]
 \centering
   \includegraphics{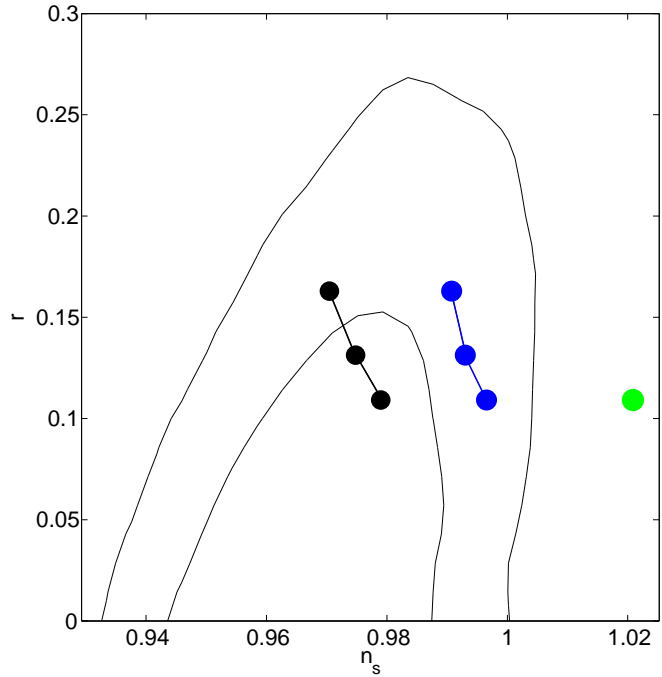}
    \caption{Same contours  in the $n_s$-$r$ plane as in Fig. 5. The
    theoretical values for the case of the GB patch correspond to
    $p=2$ (dots) with $N$=50, 60 and 70 (top to bottom). Two  of (non)commutative
    inflation are at left ($\tilde{\sigma}=0$) and center
    ($\tilde{\sigma}=2$). The case of $\tilde{\sigma}=6,N=70$ appears at right.
     Note that $p=4$ is not present at this graph.}
    \label{fig6}

\end{figure}

\begin{figure}[t!]
 \centering
   \includegraphics{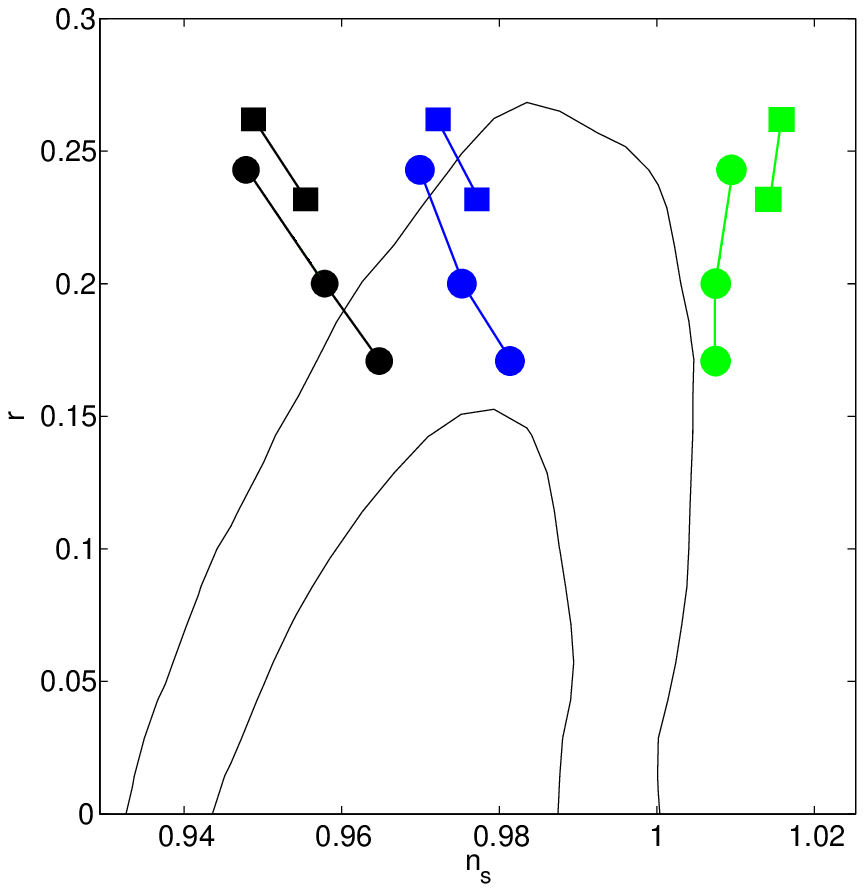}
    \caption{Same contours in the $n_s$-$r$ plane as in Fig. 5. The
    theoretical values for the case of the RS patch correspond to
    (a) $p=2$ (dots) with $N$=50, 60 and 70 (top to bottom).   (b) $p=4$
    (squares) with $N=60$ and 70.  Three classes of (non)commutative
    inflation are at left ($\tilde{\sigma}=0$), center
    ($\tilde{\sigma}=2$), and right ($\tilde{\sigma}=6$).} \label{fig7}
\end{figure}

For large-field potentials, the spectral index $n_s$ and
tensor-scalar ratio $r$ depend on the e-folding number $N$ only.
This simplicity leads to strong constraints on large-field
potentials. Further, combining the Gauss-Bonnet braneworld with
large-field potentials provides even tighter constraints than for
the GR case. It was shown with WMAP1 that the quartic potential
$V=V_{04} \phi^4$ is under strong observation pressure (ruled out
observationally) for GR and RS (GB) patches, while the quadratic
potential $V=V_{02}\phi^2$ is inside the 1$\sigma$ bound for GR
and GB patches for a range of e-folding number $50\le N\le 60$.
This potential, however, is outside the 1$\sigma$ bound for RS
patch.  This is obtained from the likelihood analysis of $n_s$ and
$r$ with the leading-order calculation for patch cosmological
models~\cite{CT}.

The main feature of the combined DATA is the red-tilted scalar
spectral index $n_s<1$ and the small tensor-scalar ratio $r<0.22$.
We show the theoretical values of $n_s$ and $r$ for $N=50$ and 60
in Table 2. For $\tilde{\sigma}=2$, $p=4$ in GB and all cases of
$\tilde{\sigma}=6$, we have blue-tilted scalar spectra with
$n_s>1$. Also, in all cases $p=4$ leads to $r>0.22$, which is
beyond the $2\sigma$ bound. In order to compare these values with
the combined DATA, we should have nine $n_s$-$r$ figures with
different contours, in order to take into account all possible
combinations of $q$ and $\tilde{\sigma}$.  As was shown in the
figures constructed by the WMAP1~\cite{CT}, however, these
contours are similar except for a minor modification of the upper
bound on $r$. Hence, we use Fig. 5 constructed for the
$\tilde{\sigma}=0$ GR patch with the combined DATA~\cite{SSM} to
plot all theoretical values of large-field potentials, instead of
generating different contours for each combination of $q$ and
$\tilde{\sigma}$.

We start with the GR patch with $\tilde{\sigma}=0$. The
large-field potential $V \propto \phi^p$ is consistent with the
combined DATA for $p=2$, and is marginally consistent with the
WMAP3 alone for $p=4$ but ruled out by the WMAP3+SDSS~\cite{KKMR}.
In Fig. 5, we see that the quadratic potential is inside the
$2\sigma$ contour, while the quartic potential is outside the
$2\sigma$ contour for $50 \le N \le 70$\footnote{Here we include
$N=70$ because this number occurs naturally in the brane world
models~\cite{SSS}.}. Hence we find that the quartic potential is
ruled out by the combined DATA~\cite{SSM}. The weak noncommutative
case of $\tilde{\sigma}=2$ is favored for the observational
compatibility of the quadratic potential, while the strong
noncommutative case of $\tilde{\sigma}=6$ is disfavored. The
quartic potential is marginally compatible for $\tilde{\sigma}=2$
and it is not allowed for  $\tilde{\sigma}=6$. The authors of
\cite{CT} argued that the quartic potential is rescued from the
marginal rejection in the $\tilde{\sigma}=2$ GR case when using
the WMAP1 data. However, it is not clear whether this is correct,
since this case is on the border of  the $2\sigma$ contour.

Now we are more on to the GB patch. When combined with the
Gauss-Bonnet braneworld, it was recently shown that the GB patch
may provide a successful cosmology~\cite{Pan}. The GB and GR
patches provide nearly the same result for the quadratic potential
according to the WMAP1 ~\cite{CT}. In Fig. 6 we see that the
quartic potential is ruled out observationally for all $
\tilde{\sigma}=0$, $2$ and $6$, while the quadratic potential is
located near the $1\sigma$ bound for $\tilde{\sigma}=0$ and inside
the $2\sigma$ bound for $\tilde{\sigma}=2$. However, it is outside
the $2\sigma$ bound for the strong noncommutative case with
$\tilde{\sigma}=6$.  Hence the quadratic potential in GB patch is
consistent with the combined DATA, as is similar to the GR case.
At this stage we compare the RS patch with the combined DATA. As
is shown in Fig. 7, the quartic potential is under strong
observational pressure, similar to the GR case. The case of
$\tilde{\sigma}=6$, $p=2$ is  outside the $2\sigma$ bound, and the
$\tilde{\sigma}=0$, $p=2$ and $\tilde{\sigma}=2$, $p=2$ cases are
on the boundary of the $2\sigma$ contour.

On the other hand, we feel from Figs. 5 and 7 that the quartic
potential might become marginally compatible in $\tilde{\sigma}=2$
GR and $\tilde{\sigma}=2$ RS patches. There is no significant
difference between GR and RS patches even though the tensor-scalar
ratio $r$ is slightly larger in RS than in GR.  Then it would be
premature to claim that the quartic potential is ruled out in the
Gauss-Bonnet braneworld. In this case it is better to study the
exact Friedmann equation (\ref{Eeq}) than the effective Friedmann
equation (\ref{Heq}). As an example, one may consider the passage
of RS $\to$ GR $\to$ GB. Here the GR exists as the intermediate
regime. In this case the quartic potential in GR patch might come
within the $2\sigma$ bound ~\cite{SS}.

Finally, we wish to mention the inflation induced by tachyon
field. It is known that the tensor-scalar ratio is smaller in the
tachyon inflation than the scalar inflation irrespective of the
kind of patch cosmologies\footnote{Here we have the tensor-scalar
ratio for the tachyon inflation
$\tilde{r}^{LF}=\frac{8qp}{\zeta_q} \frac{1}{\tilde{X}}$ with
$\tilde{X}=[pq + 2]N + pq/2$~\cite{CT}.}. Hence we expect that
this leads to the compatibility with the combined WMAP3. Further,
the effect of noncommutativity could induce  a blue-tilted
spectrum with $n_s>1$ as is similar to the standard scalar field.
However, we did not investigate the tachyon inflation  because we
focused on the standard scalar inflation in this work.

In conclusion, the quadratic potential is acceptable for GR and GB
patches, while the quartic potential is ruled out by the combined
DATA in most of patches. However, there is a possibility that the
quartic potential is marginally compatible with the combined DATA
in $\tilde{\sigma}=2$ GR and $\tilde{\sigma}=2$ RS patches. We
note that the strong noncommutative inflation $\tilde{\sigma}=6$
is excluded because it leads to blue-tilted scalar spectra. A more
thorough comparison of the combined DATA with (non)commutative
patch models would necessarily involve computing all nine
likelihood contours.

\subsection*{Acknowledgments}
B. Murray was supported by the Department of Energy under
DE-FG06-85ER40224. Y. Myung was supported by the Korea Research
Foundation Grant (KRF-2005-013-C00018).

\end{document}